# DFT Investigation of Osmium Terpyridinyl Complexes as Potential Optical Limiting Materials

Shashwat Alok and Jason Gunson, *†

Department of Chemistry, University of Fribourg, Avenue de l'Europe 20, 1700 Fribourg, Switzerland

*Supporting Information Placeholder*

**ABSTRACT:** The development of optical power limiting materials is important to protect individuals or materials from intense laser irradiation. The photophysical behavior of Os(II) polypyridinyl complexes having aromatic hydrocarbon terpyridyl ligands has received considerable attention as systems exhibiting intramolecular energy transfer to yield a long excited states lifetime. Here we present a focused discussion to illustrate the photophysical behavior of transition metal complexes with modified terpyridyl ligands, utilizing density functional theory. Our DFT studies of the excited state behavior of Os(II) complexes containing pyrene-vinylene derived terpyridine (pyr-v-tpy) ligands can be applied to the development of optical limiting materials controlling the laser power at longer wavelength range.

Due to their exceptional properties in redox chemistry, spectroscopy, structural and catalytic preference, transition metal (M= Ru, Os) complexes have been extensively explored.[1] The photophysical and phtotchemical behavior of transition metal diimine and triimine complexes has been the focus of numerous investigations for more than 40 years.[2] Os(II) diimine or triimine complexes have been extensively studied for a wide range of applications. In part the development of this area is driven by the potential of diimine or triimine complexes of this metal to serve as visible light absorbing chromophores for light induced redox reactions.[2c] The visible absorption of the complexes in this class of chromophores arises from metal-to-ligand charge transfer transitions, or called MLCT.[3] In fact, the absorption, (or the color) of these complexes, can be tuned by manipulating the electronic properties, i.e. donating/withdrawing of the coordinated ligand(s) bound to the metal centers. Such tuning can be also achieved by changes in the spectator ligands. Several papers have reviewed such cases.[1a,2b,c]

In general, these complexes have the lowest energy allowed electronic transition: a metal localized d$\pi$ orbital and an imine $\pi$* orbital as acceptor.[4] A general Jablonski energy diagram, shown in Figure 1, has been used to describe the photophysical behaviors of these complexes. After excitation, the lowest energy spin allowed absorption is the singlet metal-to-ligand charge transfer ($^1$MLCT) state. This state will undergo intersystem crossing to populate either the triplet $^3$MLCT, ligand field ($^3$LF) or intra-Ligand ($^3$IL) states. The MLCT, which is an emissive state in d$^6$ Os(II) complexes, generally occurs between a reducing metal center and acceptor ligands. $^3$MCs (metal center states, LF or dd) that can quickly non-radiatively relax to return back to the ground state and are transitions localized on the metal center. $^3$IL (or ligand centered), excited states usually have a µs or ms life time and can couple to the MLCT state with the same spin multiplicity and extend the metal complex's life time.[5]

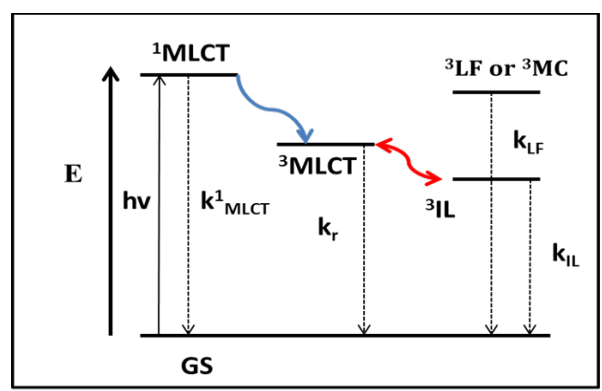

**Figure 1.** Jablonski energy diagram of Os(II) diimine or triimine complexes.

With regard to Ru(II) bisterpyridine complexes, Os(II) complexes have an advantage that the spin forbidden absorption populates the $^3$MLCT directly from the ground state and has significant absorptivity. For [Os(tpy)$_2$]$^{2+}$, the spin forbidden MLCT band is pretty intense because of the large spin-orbital coupling that results from the heavy osmium atom. Contrary to [Ru(tpy)$_2$]$^{2+}$, [Os(tpy)$_2$]$^{2+}$ shows a relative high intense luminescence at room temperature. This indicates that in the bis-terpyridine osmium complexes the short lived, non-emissive $^3$MC cannot be populated at room temperature.[2b,4] However, the disadvantage of this system is that the life time of the excited state is comparably short, which would limit the utility of this complex to protect against pulsed laser sources with longer pulse duration. The other disadvantage of the osmium complexes is their low $^3$MLCT (triplet metal to ligand change transfer state), which requires an even lower triplet $^3$IL from the ligands to coordinate. In this manuscript, we will demonstrate how to utilize DFT calculation to help rationalized designing osmium bisterpyridine complexes that could potentially be used as optical limiting materials. Time dependent DFT (TDDFT) has also been applied to corroborate the relative transitions.

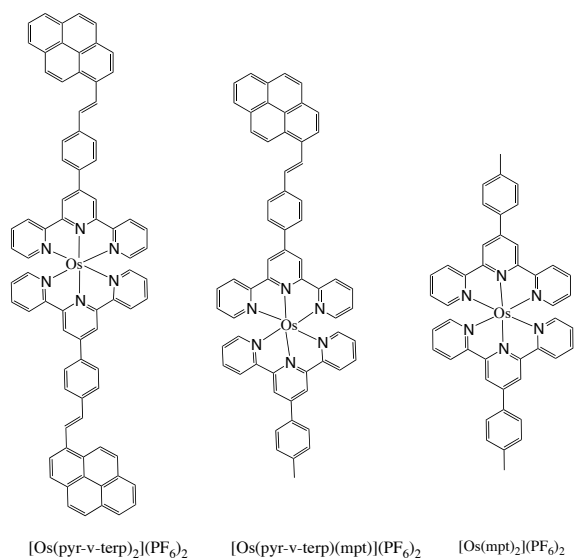

[Os(pyr-v-terp)$_2$](PF$_6$)$_2$    [Os(pyr-v-terp)(mpt)](PF$_6$)$_2$    [Os(mpt)$_2$](PF$_6$)$_2$

**Figure 2.** Structures of Osmium bis-terpydine complexes.

The continuing integration of high power light sources, such as visible, near infrared and infrared lasers system into the modern technology or used for military purposes has led to the development of an ever increasing number of novel materials including porphyrins, phthalocyanines and organometallic complexes which might serve for protecting human eyes or others. Although Os(II) diimine complexes have been extensively studied for a wide range of applications, studies of the excited state behavior of Os(II) triimine complexes containing pyrene-vinylene derived terpyridine ligands have been of fewer examples.[6] Here we focused on theoretical investigation of the Os(II) complexes, particularly the terpyridinyl compounds, i.e pyrene-vinylene derived terpyridine (py-v-tpy) ligated complexes towards elongated lifetime at excited state and extended absorption in visible range.

Several complexes as show in Figure 2 were the target molecules. The theoretical model has been previously explored successfully to simulate the Ru triimine or diimner or other complexes.[2c,7,8] The approach was achieved based on density functional theory according to previous published method as detailed in following:[9] Geometry optimizations of these complexes were performed with the Gaussian 09 suite of software and employed the B3LYP functional. The authenticity of each converged structure was confirmed by the absence of imaginary vibrational frequencies. A double-ζ (DZ) basis set with an effective electron core potential (LANL2DZ ECP) was used for Os, and basis sets [(6-311+(d,p), 6-31(d,p), 6-31(d)] were all evaluated in comparison for the remaining atoms. Orbital images were created with the use of ChemCraft, as shown in supporting information. Triplet or singlet excited states calculations were obtained. Time dependent DFT has also been applied to predict the transition and their relative absorptions. The polarizable continuum model (PCM) was applied to model solvent effects.

First, there basis sets [(6-311+(d,p), 6-311(d,p), 6-31(d)] for C, H, N were evaluated for the structures optimization. The ground states calculations for all these complexes can be achieved in reasonable computing time, however, excited states calculations utilizing 6-311+(d,p) and 6-31(d,p) can not be converged. Therefore, the discussions and results in this manuscript are based on the level of accuracy at 6-31(d). The excited state calculations are employed with their particular method as indicated in each specific case as shown in supporting information. In addition, simple amine or dien ligand was applied to substitute the terpyridine ligand in order to reasonably model the Os complexes. Such substitution can significantly save the computing time but also can successfully predict the photo-physical properties of this type complex as evidenced by Ru complexes in previous publication.[9] Moreover, the ligand's optical properties were computed when the ligand was coordinated with a fully loaded d$^{10}$ metal that is not active for any metal to ligand charge transfer, for instance, [Zn(py-v-tpy)$_2$]$^{2+}$, hence this complexes have been calculated and compared when the MLCT states has been completely excluded.

Ground state optimized structures of ligand py-v-tpy; [Os(tpy)(NH$_3$)$_3$]$^{2+}$; [Os(tpy)(dien)]$^{2+}$; [Os(mpt)$_2$]$^{2+}$; [Os(pyr-v-tpy)(dien)]$^{2+}$; [Os(pyr-v-tpy)(dien) ]$^{2+}$; [Os(pyr-v-tpy)(mpt)]$^{2+}$ and [Os(pyr-v-tpy)$_2$]$^{2+}$ were achieved by the methods mentioned above. Selective optimized structures were shown in Figure 3. The frequencies of the complexes have also been calculated, corroborating the global minimum energy state of these complexes. Excited states calculations are more complicated, and consume more computation time. Triplet-state energy approximation can be achieved via Density Functional Methods and the obtained photo-physics data can be corroborated by quenching studies, electrochemistry and transient absorptions. A key difficulty in characterizing the lowest-energy long-lived excited state of these complexes is the fact that, in many cases, the directly observation of phosphorescence for the pyr-v-tpy ligand, or other larger ligand, or either of their coordinated complexes is not achievable. As a result, we have employed hybrid density functional calculations to bridge the gap between structurally related systems with known triplet excited-state energies and the pyr-v-tpy complexes reported here. Triplet quenching studies from experimental data were obtained from reference[4] because no triplet formation was observed upon direct photolysis of the ligand; thus, the assumption is made that the triplet energy of this complex is the same as the triplet energy of the ligand. In reference [4], rate constants for quenching of the transient absorption of the complex were obtained with several quenchers at different energy states. It is noteworthy that the quenching study applied on Os complexes is quite similar to the Ru complex with same coordination ligands. The computational approach employed was to estimate the ligand ground state-lowest-energy triplet state zero-zero energies using the hybrid B3LYP functional with an intermediate basis set of 6-31G(d). The energy difference was taken as the difference in energy of the gas-phase or solution-optimized geometries of the two states. This approach provides reliable values of the triplet-state energies for the structurally related aromatic hydrocarbon references used.

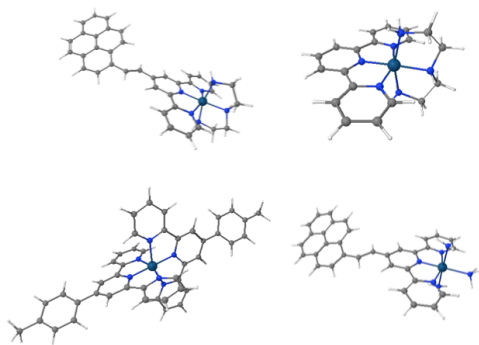

**Figure 3.** Optimized structures of Osmium complexes complexes.

Table 1 lists calculated and experimental values (from reference[4]) for two aromatic hydrocarbons as well as calculated values for pyr-v-tpy (ligand) and a Zn(II) complex containing pyrv-bpy, $[Zn(pyr-v-tpy)_2]^{2+}$. The triplet energy for the Zn(II) complex was calculated because the lowest-energy state is ligand-localized but reflects an increased degree of pyrene-to-terpyridine charge-transfer interaction following the coordination of the terpyridine ligand to Zn(II). The result is that the triplet energy of the pyrv-bpy coordinated to Zn(II) is significantly lower than the triplet energy of the free ligand.

Similar to previously published Ru(II) complexes, the Os(II) complexes can be approached by employing two methods to estimate excited-state energies. A method similar to that for the ligands was used, except that the ground state and lowest-energy triplet were calculated for the complex in the presence of solvent. The hybrid B3LYP functional was used, and the basis set was LANL2DZ for all of the atoms. Other methods with higher-level basis set can not be achieved due to extremely long time computation time, particularly on the excited states calculations. By calculating the energies of the solvated singlet and lowest-energy triplet, a measure of the zero-zero energy is obtained. An alternate approach to estimating the energy of the lowest-energy triplet was via time-dependent density functional calculations on the geometry-optimized singlet ground states of the chromophores. In this case also, the method used was B3LYP and the basis set for the light elements was 6-31G(d) whereas that for Os was LANL2DZ. Time-dependent DFT calculations for the ground-state optimized geometry included acetonitrile solvent using the polarized continuum model.

**Table 1. Calculated DFT and TDDFT data compared to experimental observations.**

| Ligand/ Complex | DFT $\Delta E(S_0-T_1)$, cm$^{-1}$ | TDDFT $\Delta E$ $(S_0-T_1)$, cm$^{-1}$ | $\Delta E$ (Exp.) cm$^{-1}$ | Transition |
|---|---|---|---|---|
| **Pyrene** | 16,900 | 16,900 | 16,900 | π-π* |
| **Pyr-v-ph** | 13,700 | 13,700 | 13,600 | π-π* |
| **Pyr-v-terp** | 14,000 | 13,900 | 13,900 | π-π* |
| **$[Zn(pyr-v-typ)_2]^{2+}$** | 14,700 | 14,600 | 14,500 | π-π* |
| **$[Os(mpt)_2]^{2+}$** | 14,900 | 15,000 | 14,900 | MLCT |
| **$[Os(per-v-tpy)(mpt)]^{2+}$** | 14,800 | 14,900 | - | MLLCT |
| **$[Os(pyr-v-terp)_2]^{2+}$** | 14,700 | 14,700 | - | MLLCT |

The result provides a measurment of the vertical transition energies for the complexes, or called ground-state singlet to lowest-energy triplet vertical transitions.[10] The ligand calculations are obviously accurate when comparing DFT or TD-DFT data with the experimental observations. For $[Zn(pyr-v-tpy)_2]^{2+}$, the lowest-energy transition is at 690 nm which perfect matched with the calculated value, 14 500 cm$^{-1}$. The orbitals of the transition were listed in these complexes are all π - π*. The lowest singlet-triplet transition in Os complexes are Os dπ to bpy π* MLCT. For this complex specifically, the CI coefficient expansion demonstrateds that the transition has contributions from MLLCT,[11] Os dπ to Pyr-v-tpy π*, and a rather diffuse dπ, Pyr-v-tpy π-to- Pyr-v-tpy π* transition.[12] The influence of the styryl pyrene substituent, as shown in table 1, from $[Os(mpt)_2]^{2+}$; $[Os(py-v-tpy)(mpt)]^{2+}$ to the complexes $[Os(py-v-tpy)_2]^{2+}$, is readily apparent in lowering the energy of the transition and in increasing the degree of ligand localization in the transition from calculaitons, although the experimental value for $[Os(py-v-tpy)(mpt)]^{2+}$ has not been available at this time.

In conclusion, the DFT or TDDFT calculations help support the concept that the lowest-energy transition on the pyr-v-tpy containing complexes is, to a significant extent, localized on the pyr-v-tpy ligand. In addition the energy gap between this lowest-energy triplet state and the higher-energy MLCT state is certainly similar to Ru complexes at the same coordination ligands. Therefore, this type of pyr-v-tpy derived Os complexes might be also a candidate for optical limiting material.

## ASSOCIATED CONTENT

### Supporting information

Optimized structures both in gas phase and in acetonitrile solvent; Excited states calculation details; details of computational methods.

## AUTHOR INFORMATION

### Corresponding Author

Jason.Gunson@gmail.com

### Present Addresses

† Swiss Academy of Sciences, Switzerland